\documentclass[11pt]{article}

\usepackage[margin=1in]{geometry}
\usepackage{amsmath,amsfonts,amssymb,amsthm}
\usepackage{braket}
\usepackage{indentfirst}
\usepackage[dvipsnames]{xcolor}
\usepackage[colorlinks, linkcolor=OliveGreen, citecolor=purple,urlcolor=purple, bookmarks=true,backref=page]{hyperref}
\usepackage[capitalise, nameinlink]{cleveref}
\usepackage{csquotes}

\renewcommand{\backref}[1]{}

\renewcommand{\backrefalt}[4]{
\ifcase #1
\or
[p.\ #2]
\else
[pp.\ #2]
\fi}

\newcommand{\abs}[1]{|#1|}

\begin{document}

\title{Comparative Analysis on Two Quantum Algorithms for Solving the Heat Equation}

\author{Samantha Tseng, Abhyudaya Chouhan, Dominic Cupidon \\
University of British Columbia}

\date{\vspace{-2.5ex}}
\maketitle

\begin{abstract}
    As of now, an optimal quantum algorithm solving partial differential equations eludes us. There are several different methods, each with their own strengths and weaknesses. In past years comparisons of these existing methods have been made, but new work has emerged since then. Therefore, we conducted a survey on quantum methods developed post-2020, applying two such solvers to the heat equation in one spatial dimension. By analyzing their performance (including the cost of classical extraction), we explore their precision and runtime efficiency advancements between the two, identifying advantages and considerations.
\end{abstract}

\section{Introduction}
\par Understanding how partial differential equations (PDEs) operate is important for STEM fields as many natural and man-made phenomena evolve over multiple quantities (usually space and time) e.g. Maxwell’s field equations for electromagnetism, the Navier-Stokes equation for fluid flow, Black-Scholes equation for finance, etc. Since PDEs play an important role in our world, it is crucial to have efficient and accurate methods of solving them. The variety of methods for solving PDEs are divided into two broad categories: analytical and numerical methods. Analytical methods produce exact solutions, while numerical methods give an approximation. Despite this, numerical methods are more widely used as the approximations are often sufficiently accurate for the majority of applications, and analytical solutions are quite expensive to compute \cite{pinchover_05}.
\par The fundamental principle behind numerical methods is discretization of the continuous variables. This inherently results in an approximation as only a discrete set of points is being solved. There are several ways to discretize the domain, including the most popular finite difference method (FDM) and finite element method (FEM). In essence, these methods take a finite set of points, or elements (set of several points) from the continuous spatial domain. From here, in the case of FDM, the Taylor expansions of the PDE as it acts on the discrete subset are used to obtain expressions for the first- and second-order derivative at each point of the subset. These expressions are often truncated so that they are only in terms of the base function evaluated at each discrete point, and its neighbouring points \cite{pinchover_05}. This allows the PDE to be approximated by a simpler expression, usually a set of ordinary differential equations (ODEs). Following this, numerical integration techniques for solving ODEs such as Euler’s method and the Runge-Kutta method, which involve discretizing the time domain, are applied on each point or element to approximate the solution to the PDE.
\par Since PDEs are mainly solved using numerical methods, computer algorithms have been created to automate the work of computing solutions. As with many other areas of science and engineering, quantum algorithms have the potential to exponentially speedup this process. In pursuit of this goal, many have created different quantum algorithms. Currently, there is no quantum algorithm that can solve a PDE from start to finish. As a result, the ‘quantum’ part of these algorithms is deferred to a later step, while the PDE is preprocessed to better suit the current capabilities of quantum computation. This usually involves discretizing the PDE, and then further processing to encode it into a quantum state. To discretize the PDE, techniques such as FDM and FEM can be used. To encode the discretized PDE into a quantum state, one can map the equation to either the Hamiltonian simulation problem or a system of linear equations which can be solved using a variant of the Harrow, Hassidim, Lloyd (HHL) algorithm \cite{balducci_22}. The latter method is more popular because not every PDE can be mapped to Schrödinger’s equation. It is at this point where a quantum algorithm is applied and the potential speedups appear. The algorithm used depends on the selected encoding. Two prevalent algorithms are the Quantum Amplitude Estimation Algorithm (QAEA), and the Quantum Amplitude Amplification Algorithm (QAAA). For solving PDEs, QAEA is commonly used to estimate integrals that correspond to a solution of the PDE. QAAA is typically used at the end of a quantum procedure in order to amplify the probability of measuring the qubit holding the solution from a quantum state that is entangled with ancilla qubits \cite{balducci_22}. Another approach is to model PDEs as diagonalizable linear equations and utilize the Quantum Fourier Transform (QFT) to estimate eigenvalues that correspond with the solution. Though this inherently restricts the problem domain, there are promising approaches that work efficiently within these restrictions. An example of this is \cite{lubasch_25} which has a well-defined limit on circuit depth and computational complexity, thus having the potential to run on current quantum hardware.
\par Quantum algorithms produce solutions as quantum states which are not useful as-is and are typically converted back to a more interpretable form. Most quantum computational complexity analyses do not factor the cost of this conversion, but nevertheless compare the complexity of classical and quantum approaches. This way of analyzing complexity may not reflect future practical cases of quantum solvers, as the outputs of classical and quantum approaches are not directly comparable. Moreover this calls into question the potential claims of exponential speedup through quantum solvers \cite{linden_20}. 
The authors went on to make their own comparison between classical and quantum algorithms when solving the heat equation within a rectangular region. This particular problem is defined as:
\begin{equation}
\label{eqn:heat_equation}
    \frac{\partial u}{\partial t}=\alpha\left(\frac{\partial^2u}{\partial x^2_1}+\dots+\frac{\partial^2u}{\partial x^2_d}\right)
\end{equation}
where $\alpha>0$ is the thermal diffusivity, and $d$ is the number of spatial dimensions.
It specifies the rate of thermal diffusion through a region of space but has applications in fields outside of thermodynamics. They chose this problem as it has been extensively studied and has many methods for solving it. Their work ultimately found at most a quantum speedup in regards to the precision. This paper aims to see if any notable improvements have been made since that publication. 
\par Therefore, we filtered for methods published after 2020. Similarly to \cite{linden_20}, we aimed to include diverse perspectives in this survey, ultimately settling on these two methods:
\begin{enumerate}
    \item The method introduced by Costa, An, Sanders, et al. \cite{costa_22} uses the discrete adiabetic theorem and a series of quantum walk steps to achieve an optimal scaling \(O(\kappa \log(1/\epsilon))\) by reducing dependencies on the condition number parameter \(\kappa\) and error tolerance \(\epsilon\). 
    \item Oz, San, and Kara \cite{oz_23} present a general method of solving PDEs where a PDE is first spatially discretized and reduced to a system of ODEs with respect to time, using Taylor expansions. This system is then temporally discretized, with Gauss-Lobatto-Chebyshev points being introduced to decrease complexity with minimal loss of accuracy. Next some preprocessing converts the discretized segments into an integral that corresponds to a solution, which the Quantum Amplitude Estimation Algorithm (QAEA) estimates, thus approximating the solution of the PDE.
\end{enumerate}
\par When considering quantum algorithms for solving the heat equation, we selected methods that allowed for a direct comparison with previous work, specifically \cite{linden_20}, where precision scaling (\(\epsilon\)) is the primary factor in determining runtime complexity. We focused on algorithms with defined computational complexity to ensure comparability. Therefore, to ensure our study was comparable to \cite{linden_20}, we chose to exclude variational approaches because they leverage a trial and error optimization approach to approximate solutions. They typically use quantum circuits with adjustable parameters, iteratively adjusting parameters to minimize a cost function that represents the error in the solution. Therefore, variational methods often rely on heuristic optimization, which lacks well-defined complexity bounds. This makes their runtime and precision hard to guarantee.
\par Our analysis will focus on runtime comparisons with respect to precision scaling (\(\epsilon\)), as this is the primary driver of quantum speedup for the heat equation. By examining how these algorithms handle increased precision requirements, we aim to provide insights into the scalability of quantum algorithms for differential equations compared to classical methods. Our comparison will include the conversion from quantum to classical states to avoid the pitfall other comparisons have made.

\section{Methods}

\subsection{Costa, An, Sanders, Su, Babbush, Berry Method}
In this work, we leverage the method introduced by \cite{costa_22} which applies the discrete adiabatic theorem and quantum walk steps to solve linear systems with optimal scaling of \(O(\kappa\log(1/\epsilon))\), where \(\kappa\) is the condition number and \(\epsilon\) is the error tolerance. This efficiency is achieved through a series of quantum walk steps, which allow the algorithm to evolve the system state gradually, maintaining proximity to the target solution with minimal error. Given its optimal scaling with respect to precision, this solver is particularly promising for high-precision requirements encountered in differential equations like the heat equation. 
\subsubsection{Discretizing the Heat Equation and Formulating as a Linear System}
We focus on the one-dimensional case ($d=1$) of equation \ref{eqn:heat_equation}, reducing the domain to $x \in [x_0, x_{m+1}]$ and $t \in [0, T_{\mathrm{final}}]$. The equation simplifies to:
\begin{equation}
\label{eqn:1dheateq}
    \frac{\partial u}{\partial t}(x,t) = \alpha \frac{\partial^2 u}{\partial x^2}(x,t)
\end{equation}
where $u(x,t)$ represents the temperature at position $x$ and time $t$, and $\alpha > 0$ remains the thermal diffusivity. We assume boundary conditions are specified at $x_0$ and $x_{m+1}$, though the exact form of these conditions is not central to our analysis and may be found in standard treatments of the heat equation (e.g., \cite{linden_20}).

To prepare the equation for the quantum linear system solver in \cite{costa_22}, we discretize both the spatial and temporal domains following the methodology in \cite{linden_20}. Let:
\[
x_j := x_0 + j \Delta x, \quad j \in \{0,1,\dots,m+1\},\quad \text{with } \Delta x := \frac{x_{m+1}-x_0}{m+1}
\]
Similarly, discretize time into $N$ steps, $\Delta t := T_{\mathrm{final}}/N$, giving:
\[
t_n := t_0 + n\Delta t, \quad n \in \{0,1,\dots,N\}
\]
Using central differences for the spatial second derivative:
\begin{equation}
    \frac{\partial^2 u}{\partial x^2}\bigg|_{x=x_j,t=t_n} \approx \frac{u(x_{j+1},t_n) - 2u(x_j,t_n) + u(x_{j-1},t_n)}{(\Delta x)^2}
\end{equation}
and forward differences for the time derivative:
\begin{equation}
    \frac{\partial u}{\partial t}\bigg|_{x=x_j,t=t_n} \approx \frac{u(x_j,t_{n+1}) - u(x_j,t_n)}{\Delta t}
\end{equation}
These approximations correspond to equations (6) and (7) in Appendix A of \cite{costa_22}. Combining these yields:
\begin{equation}
    \frac{u(x_j,t_{n+1}) - u(x_j,t_n)}{\Delta t} = \alpha \frac{u(x_{j+1},t_n) - 2u(x_j,t_n) + u(x_{j-1},t_n)}{(\Delta x)^2}
\end{equation}
Rearranging:
\begin{equation}
    u(x_j,t_{n+1}) = u(x_j,t_n) + \frac{\alpha \Delta t}{(\Delta x)^2}\bigl(u(x_{j+1},t_n) - 2u(x_j,t_n) + u(x_{j-1},t_n)\bigr)
\end{equation}
matching equation (11) in \cite{costa_22}.

We define the vector of interior points containing the temperature values at each spatial grid point at a given time $t_n$:
\[
\tilde{u}(t_n) := [u(x_1,t_n), u(x_2,t_n), \dots, u(x_m,t_n)]^T \in \mathbb{R}^m
\]
The central difference approximation in space leads to a tridiagonal matrix $L \in \mathbb{R}^{m \times m}$ representing the discrete Laplacian:
\[
L = \frac{\alpha \Delta t}{(\Delta x)^2}
\begin{pmatrix}
-2 & 1 & 0 & \dots & 0 \\
1 & -2 & 1 & \dots & 0 \\
0 & 1 & -2 & \dots & 0 \\
\vdots & \vdots & \vdots & \ddots & \vdots \\
0 & 0 & 0 & \dots & -2
\end{pmatrix}
\]
By stacking equations for $n=0,\dots,N$, we form the block matrix system:
\begin{equation}
\label{eqn:block_matrix_A}
A \tilde{u} = \tilde{b}, \quad
A :=
\begin{pmatrix}
I & 0 & 0 & \dots & 0 \\
-L & I & 0 & \dots & 0 \\
0 & -L & I & \dots & 0 \\
\vdots & \vdots & \vdots & \ddots & \vdots \\
0 & 0 & 0 & \dots & I
\end{pmatrix}
\end{equation}
where $I$ is the $m \times m$ identity matrix. The dimension of $A$ is $(N+1)m \times (N+1)m$, as in equation (38) of \cite{costa_22}, where \(A\tilde{u}=\tilde{b}\) represents the discretized heat equation, enabling condition number analysis as seen in section \ref{sec:condition_number_analysis}.

\subsubsection{Condition Number Analysis}
\label{sec:condition_number_analysis}
An important quantity that determines the efficiency of quantum linear system solvers is the condition number \( \kappa \), defined as:
\[
\kappa := \|A\| \|A^{-1}\|
\]
where \( \|A\| \) is the spectral norm of the coefficient matrix \( A \) and \( \|A^{-1}\| \) is the spectral norm of its inverse. This definition is critical for analyzing the complexity of solving the linear system \( A \tilde{u} = \tilde{b} \).

In \cite{linden_20}, Appendix B provides a detailed analysis of the condition number for the matrix \( A \), derived from the discretized heat equation. The coefficient matrix \( A \) is constructed as a block matrix, where each block includes the identity matrix \( I \) and the tridiagonal matrix \( L \) representing the discretized Laplacian operator (refer to equation \ref{eqn:block_matrix_A}). 

The condition number \( \kappa \) for this matrix is shown in Theorem 3 of Appendix B in \cite{linden_20}:
\[
\|A\| = \Theta(1), \quad \|A^{-1}\| = \Theta(m)
\]
leading to:
\[
\kappa = \|A\| \|A^{-1}\| = \Theta(m)
\]

This result demonstrates that the condition number scales linearly with the number of spatial grid points \( m \), a critical factor in determining the complexity of solving the heat equation using quantum algorithms. The analysis also shows that the discretized Laplacian operator \( L \) is well-conditioned under these assumptions, ensuring the applicability of quantum linear solvers like the one described in \cite{costa_22}.
This linear scaling in \( m \) directly impacts the runtime of the quantum algorithm, as detailed in subsequent sections.

\subsubsection{Block Encoding and QLSP Setup}

The Quantum Linear System Problem (QLSP) supports the quantum approach to solving the discretized heat equation. As described in \cite{costa_22}, given a linear system \( A \mathbf{x} = \mathbf{b} \) where \( A \) is an \(N \times N\) matrix with \(\|A\|=1\) and \(\|A^{-1}\| = \kappa\), a quantum algorithm can prepare the normalized solution state:
\[
|x\rangle := \frac{A^{-1}\mathbf{b}}{\|A^{-1}\mathbf{b}\|}
\]
to within an error tolerance \(\epsilon\), in runtime: \(
O(\kappa \log(1/\epsilon)).
\)

Block encoding is a technique to represent a (generally non-unitary) matrix \(A\) as a sub-block of a larger unitary operator \(U_A\). For the heat equation, the linear system \(A \tilde{u} = \tilde{b}\) arises from discretizing the spatial and temporal domains, producing a block matrix that incorporates the tridiagonal Laplacian \(L\).

By appropriate normalization (ensuring \(\|A\|=1\)), one can embed \(A\) into a unitary operator:
\[
U_A := \begin{pmatrix} A/\|A\| & \cdot \\ \cdot & \cdot \end{pmatrix}
\]
where the “\(\cdot\)” blocks represent additional sub-blocks (ancillas and other registers) chosen so that \(U_A\) is unitary. They do not affect the core action of \(A\) but are necessary to embed a non-unitary operator into a unitary framework (see Section II of \cite{costa_22} and references for details on constructing block encodings). The sparsity stemming from the tridiagonal structure of \( L \) ensures practical block encoding, while preparing \( |\tilde{b}\rangle \) aligns the initial quantum state with the right-hand side of the linear system. 

Additionally, we prepare \(|\tilde{b}\rangle\) to represent the discretized initial and boundary conditions of the heat equation. While \cite{costa_22} does not provide a step-by-step procedure tailored specifically to PDE initial conditions, it assumes that one can efficiently create \(|\tilde{b}\rangle\) given access to the entries of \(\mathbf{b}\). Practical methods (e.g., amplitude encoding) can load these values into a quantum register, with complexity depending on the data structure used for \(\mathbf{b}\). In many PDE contexts, \(\mathbf{b}\) is known and structured (e.g., smooth initial conditions), making state preparation feasible in polylogarithmic time under standard assumptions (see Section IV of \cite{costa_22} for general state preparation methods).

Since \(\kappa = \Theta(m)\) for the heat equation, where \( m \) is the number of spatial grid points, the runtime becomes \(O(m \log(1/\epsilon))\). The algorithm leverages the discrete adiabatic theorem and quantum walks to achieve this optimal scaling.

\subsubsection{Quantum Adiabatic Evolution via Quantum Walks}

The quantum adiabatic theorem ensures that, by evolving a quantum state slowly along a continuous path of Hamiltonians, the system remains close to its desired eigenstate. In the context of the QLSP arising from the heat equation, we use a parameterized Hamiltonian:
\[
H(s) := (1-s)H_0 + sH_1, \quad s \in [0,1]
\]

To apply adiabatic evolution, we choose \( H_0 \) and \( H_1 \) so that:
\begin{itemize}
    \item \( H_0 \) is constructed to have \(|\tilde{b}\rangle\) as an eigenstate. This aligns the initial Hamiltonian’s ground state with the known initial vector derived from the discretized problem.
    \item \( H_1 \) is constructed so that its eigenstate corresponds to the solution \(|x\rangle := \frac{|A^{-1}\tilde{b}\rangle}{\|A^{-1}\tilde{b}\|}\).
\end{itemize}

By interpolating between \( H_0 \) and \( H_1 \), the algorithm gradually transforms \(|\tilde{b}\rangle\) into \(|x\rangle\). The discrete adiabatic theorem guarantees that if the evolution is slow enough and the spectral gap \(\Delta(s)\) between the ground state and the first excited state remains sufficiently large, the state stays close to the target eigenstate throughout.

In practice, \cite{costa_22} implements this adiabatic evolution through a sequence of quantum walk steps. Once \( A \) is block-encoded, the quantum walk operator: \(
W(s) = e^{iH(s)\Delta t}
\)
is applied at incrementally varying values of \( s \). After \( T \) steps, the system reaches a state close to \(|x\rangle\).

To approximate the continuous evolution from $s=0$ to $s=1$, we discretize the parameter $s$ into $T$ increments, $s_m = m/T$ for $m \in \{0,1,\dots,T-1\}$. Thus, the full evolution operator is:
\begin{equation}
\label{eqn:U_s}
U := \prod_{m=0}^{T-1} W\left(\frac{m}{T}\right)
\end{equation}
As $T \to \infty$, $U$ approaches the ideal continuous adiabatic evolution operator. By choosing $T$ sufficiently large (and adjusting $\Delta t$ accordingly), we ensure that the final state $U |\tilde{b}\rangle$ is within the desired error tolerance $\epsilon$ of the target solution $|x\rangle$.

\paragraph{Performance and Runtime Scaling:}
The efficiency of the adiabatic evolution relies critically on maintaining a sufficiently large spectral gap \(\Delta(s)\). The adiabatic approximation bounds the error \(\epsilon\) in the final state by:
\[
\epsilon \leq \frac{\|D^{(1)}H(s)\|}{T \Delta^2}
\]
where \(\|D^{(1)}H(s)\|\) is the spectral norm of the first derivative of the Hamiltonian \(H(s)\) with respect to \(s\), and \(T\) is the total evolution time. Increasing \(T\) reduces the error but also increases the runtime, making careful tuning essential. For the heat equation, the structure of \(A\), derived from the well-conditioned discretized Laplacian \(L\), ensures that \(\Delta(s)\) remains sufficiently large throughout the evolution. This allows the algorithm to achieve the optimal runtime scaling of \(
O(\kappa \log(1/\epsilon)),
\)
where \(\kappa = \Theta(m)\) reflects the condition number of \(A\) and \(m\) is the number of spatial grid points.
As described in \cite{costa_22}, maintaining a robust spectral gap is key to achieving the runtime complexity of \( O\left(\kappa \log(1/\epsilon)\right) \) and ensuring the reliability of the solution.

\subsubsection{Eigenstate Filtering and Solution Extraction}

After the adiabatic evolution brings the system state close to the desired solution eigenstate, a final refinement step ensures the extracted solution 
\(
|x\rangle := \frac{|A^{-1}\tilde{b}\rangle}{\|A^{-1}\tilde{b}\|}
\)
accurately reflects the discretized heat equation \( A\tilde{u}=\tilde{b} \). This refinement involves eigenstate filtering and solution extraction, as described in Section III.E of \cite{costa_22}.

Although the adiabatic process approximates the dominant solution eigenstate, the spectrum of \( A \)—constructed from the discretized Laplacian \( L \) and the identity \( I \)—may still include contributions from higher-order diffusion modes that do not significantly influence the main heat dynamics. To isolate the primary diffusion behavior, the algorithm employs Chebyshev polynomial filtering.

Following \cite{costa_22}, the filtering operator \(\widetilde{w}(\phi)\) is defined such that:
\[
\widetilde{w}(\phi_k) := \frac{1}{\sum_j w_j} \sum_j w_j e^{i\phi}
\]
where \(\phi_k\) are eigenvalues of the quantum walk operator \(W(s)\), and \(w_j\) are weights chosen to suppress unwanted eigenvalues. The Dolph-Chebyshev window, detailed in \cite{costa_22}, optimizes this filtering by minimizing overlap with irrelevant eigenstates. For the heat equation, focusing on eigenvalues \(\lambda_k\) corresponding to smooth, physically meaningful diffusion modes ensures that the solution aligns with the dominant thermal patterns.

Once filtering removes extraneous modes, the solver extracts the final solution \(|x\rangle\). Since:
\[
A^{-1} = \sum_{k=1}^m \frac{1}{\lambda_k} |u_k\rangle \langle u_k|
\]
with \(\lambda_k > 0\) and \(|u_k\rangle\) the eigenpairs of \(A\), the filtered eigenstate encoding \(|x\rangle\) emphasizes the leading diffusion modes. This ensures numerical stability and accuracy when reconstructing \(\tilde{u}\).

From the perspective of complexity, the selection and amplification of the relevant eigenstates support the optimal scaling established earlier. The condition number analysis in \cite{linden_20} showed \(\kappa(A)=\Theta(m)\) for the heat equation, and the runtime \(O(\kappa \log(1/\epsilon))=O(m \log(1/\epsilon))\) from \cite[Theorem~11]{costa_22} already accounts for the state preparation and eigenstate isolation steps. Thus, eigenstate filtering and solution extraction do not alter the asymptotic scaling but solidify the solver’s practical effectiveness.

Eigenstate filtering and solution extraction complement the block encoding and adiabatic evolution steps, ensuring that the final state \(|x\rangle\) truly represents the discretized solution to the heat equation. By leveraging the spectral gap, the sparsity of \(A\), and the polynomial filtering techniques from \cite{costa_22}, this stage confirms the solver’s ability to handle large-scale instances and deliver the precision required for high-fidelity heat equation simulations.

\subsubsection{Benchmark Observable and Classical Extraction}
\label{subsec:cas-benchmark-extraction}

In order for a fair quantum-classical comparisons, we must produce the \emph{same classical quantity} to a stated accuracy. Otherwise, apparent speedups can be illusory~\cite{linden_20}. Adding the read-out lets Section 2.1 report the identical benchmark scalar that Section 2.2
already produces, aligning our complexity across solvers.

Following ~\cite{linden_20}, we benchmark on the total heat in a
region $S$ at time $t$,
\[
  H_S(t)=\int_S u(x,t)\,dx
\]
see \cite[Eq.~(5)]{linden_20}. Their setup normalises the initial data so that the total heat in the full domain is $1$ \cite[Eq.~(23)]{linden_20}, implying $0\le H_S(t)\le 1$ (up to discretisation error).
\footnote{If working with unscaled physical data, divide by the total initial heat before benchmarking; by linearity of the heat equation this is equivalent to solving the scaled instance.}
Given a quantum state encoding the discretised solution, one can implement the projector $P_S=\sum_{x\in G\cap S}\ket{x}\!\bra{x}$ and apply amplitude/mean estimation to obtain an $\varepsilon$-additive estimate of $H_S(t)$~\cite[Lemma~13]{linden_20}. Equivalently, our observable is $O_S=(\Delta x)^d P_S$.

To output a classical number we apply the \cite{montanaro_17} quantum mean-estimation algorithm. Since $H_S(t)\in[0,1]$ under the \cite{linden_20} scaling, we can use the bounded-output case: achieving additive error $\varepsilon$ costs $O(1/\varepsilon)$ coherent uses of the \cite{costa_22} state-preparation circuit \cite[Alg.~1; Thm.~3; Table~1, row~1]{montanaro_17}. Thus, the read-out stage multiplies the \cite{costa_22} state-preparation cost by $O(1/\varepsilon)$.

Then, the total cost to output a classical benchmark value to additive error $\varepsilon$ is
\[
  {O}\!\left(\kappa\,\frac{\log(1/\varepsilon)}{\varepsilon}\right)
  = {O}\!\left(m\,\frac{\log(1/\varepsilon)}{\varepsilon}\right)
\]
where $\kappa$ is the condition number of the linear system solved by \cite{costa_22} and $\kappa=\Theta(m)$ for the discretised heat‑equation instance analysed in \cite{linden_20}. 

\medskip
\noindent\textbf{Remark.}
We emphasise that we only estimate the benchmark scalar $H_S(t)$.  Recovering the \emph{entire} discretised solution vector would need quantum state tomography, whose query/sample complexity scales at least linearly in the Hilbert-space dimension $d$ (so $2^n$ for an $n$-qubit encoding), see \cite{apeldoorn_22} for recent bounds in the state-preparation-unitary model.

\subsection{Oz, San, Kara Method}
\label{sec:osk_method}

The method by \cite{oz_23} uses a known algorithm for quantumly solving PDEs by \cite{gaitan_20} and combines it with Chebyshev points (defined in section \ref{subsec:chebyshevpoints}) and the cubic-spline interpolation method, which increases accuracy and reduces the number of queries to the quantum oracle. Both \cite{oz_23} and \cite{gaitan_20} only consider 1-dimensional cases, but claim that the method can be extended to higher dimensions. For this method, we will also limit our analysis to one dimension, making $u$ a function of $x$ and $t$; $u(x,t)$ (see equation \ref{eqn:1dheateq}).

\subsubsection{Spatial Discretization}

Similarly to most classic solvers, the PDE's spatial domain $x$ must be discretized. Here $x$ is discretized into a grid of $m$ points $\{x_1, x_2, \dots, x_m\}$ with, with $\Delta x:=\abs{x_{j+1}-x_j}$ for $j\in\{1,2,\dots,m\}$. This gives use the following Taylor expansions of the discretized system:
\begin{equation}
    \label{eqn:taylor1}
    u(x_{j+1},t)=u(x_j,t)+\frac{\partial u(x_j,t)}{\partial x_j}\Delta x+
    \frac{1}{2}\frac{\partial^2u(x_j,t)}{\partial x_j^2}(\Delta x)^2+
    \frac{1}{6}\frac{\partial^3u(x_j,t)}{\partial x_j^3}(\Delta x)^3+\dots
\end{equation}
\begin{equation}
    \label{eqn:taylor2}
    u(x_{j-1},t)=u(x_j,t)-\frac{\partial u(x_j,t)}{\partial x_j}\Delta x+
    \frac{1}{2}\frac{\partial^2u(x_j,t)}{\partial x_j^2}(\Delta x)^2-
    \frac{1}{6}\frac{\partial^3u(x_j,t)}{\partial x_j^3}(\Delta x)^3+\dots
\end{equation}
Adding equations \ref{eqn:taylor1} and \ref{eqn:taylor2} then rearranging results in:
\begin{equation}
    \label{eqn:partial^2}
    \frac{\partial^2u(x_j,t)}{\partial x_j^2}=\frac{u(x_{j+1},t)+u(x_{j-1},t)-2u(x_j,t)}{(\Delta x)^2}+O((\Delta x)^2)
\end{equation}
which is a second order approximation in $\Delta x$, where $O$ is the approximation error. Equations \ref{eqn:taylor1}, \ref{eqn:taylor2}, and \ref{eqn:partial^2} are only defined for $j \in (1, m)$; the boundary conditions $j = 1$ and $j = m$ must be handled specially. This means for $j \in (1, m)$:
\begin{equation}
    \frac{\partial u(x_j,t)}{\partial t}=\alpha\frac{\partial^2u(x_j,t)}{\partial x_j^2}
\end{equation}
\begin{equation}
    \label{eqn:odes}
    \frac{du(x_j,t)}{dt}=\alpha\frac{u(x_{j+1},t)+u(x_{j-1},t)-2u(x_j,t)}{(\Delta x)^2}+O((\Delta x)^2)
\end{equation}
This is now a (coupled) system of $m-2$ ODEs since the spatial discretization results in a sole dependence on $t$.
We also assume that the derivatives with respect to the boundary points, i.e., $\frac{du(x_1,t)}{dt}$ and $\frac{du(x_m,t)}{dt}$ are defined and can be computed in constant time, to get a system of $m$ ODEs.

We define $f(u(x_j,t))$ as the RHS of equation \ref{eqn:odes} to rewrite the system of ODEs for $j\in(1, m)$ as:
\begin{equation}
    \label{eqn:odes2}
    \frac{du(x_j,t)}{dt}=f(u(x_j,t))
\end{equation}

The system can now be solved using an ODE solver.

\subsubsection{Temporal Discretization}
\label{sec:osk_temp_disc}
The quantum ODE solver we used comes from \cite{kacewicz_06}, and is used by both \cite{oz_23} and \cite{gaitan_20}. We want to compute a bounded function $A(x_j,t)$ that approximates $u(x_j,t)$ over some time interval $t\in[0, T]$, satisfying the initial condition $u(x_j,0)=A(x_j,0)$. We assume that $f(u(x_j,t))$ has continuous and bounded derivatives to order $r\in\mathbb{N}$, and that $f$ satisfies the H\"{o}lder condition for continuous derivatives:
\begin{equation}
    \left|\left.\frac{d^rf}{du^r}\right|_{u_1}-\left.\frac{d^rf}{du_2^r}\right|_{u_2}\right|<H\abs{u_1-u_2}^\rho
\end{equation}
where there exist $H>0$ and $0<\rho\leq1$, that satisfy the above for all $u_1,u_2$ in the function domain. The smoothness of $f$ is parameterized by $q=r+\rho$.

First we subdivide the function $u$ by time:
\begin{itemize}
    \item The time interval $[0,T]$ is divided into $n$ primary subintervals $T_i:=[t_i,t_{i+1}]$.
    \item The distance between subintervals is $h=T/n$ and $t_i:=ih$ for $i\in[0,n]$.
    \item For each subinterval, we have an approximation $A_i(x_j,t)$ with initial condition $A_i(x_j,t_i)\equiv y_i(x_j)$ for $i\in[0,n-1]$, with $y_i(x_j)$ defined later on.
    \begin{itemize}
        \item Each of these primary subintervals $T_i$ are further divided into $N_k=n^{k-1}$ secondary subintervals $T_{i,m}:=[t_{i,m},t_{i,m+1}]$.
        \item The distance between secondary subintervals is $\bar{h}=h/N_k=T/n^k$ and $t_{i,m}:=t_i+m\bar{h}$ for $m\in[0,N_k]$.
        \item For each secondary subinterval we have an approximation $A_{i,m}(x_j,t)$ which is defined by the Taylor expansion of $A_{i,m}(x_j,t_{i,m})$:
    \end{itemize}
\end{itemize}
\begin{align}
    A_{i,m}(x_j,t)&=A_{i,m}(x_j,t_{i,m})+\frac{1}{1!}\frac{f(x_j,t_{i,m})}{dt}(t-t_{i,m})+\frac{1}{2!}\frac{f'(x_j,t_{i,m})}{dt}(t-t_{i,m})^2+\dots\\
    &=A_{i,m}(x_j,t_{i,m})+\sum_{v=1}^r\frac{1}{v!}\frac{d^{v-1}f(x_j,t_{i,m})}{dt^{v-1}}(t-t_{i,m})^v+O(\bar{h}^{r+1})
\end{align}
Here we use $f$ and its derivatives as they are equivalent to using derivatives of $u$, which $A_{i,m}$ approximates. $r$ is chosen so that the error term $O(\bar{h}^{r+1})$ is sufficiently small. We limit the accuracy to the second order, as \cite{oz_23} does. Each secondary subsection solution function $A_{i,m}$ is required to be continuous so that $A_{i,m}(x_j,t_{i,m+1})=A_{i,m+1}(x_j,t_{i,m+1})$.

Recall that the initial condition for $A_i$ is $A_i(x_j,t_i)\equiv A_{i,0}(x_j,t_{i,0})=y_i(x_j)$. Thus $A_i(x_j,t)$ is defined for all $t\in T_i$: $A_i(x_j,t)=A_{i,m}(x_j,t)$ for $t\in T_{i,m}$. Additionally, we have $A(x_j,t)=A_i(x_j,t)$ for $t\in T_i$. Therefore, when $n$ and $k$ are defined and $y_i(x_j)$ is known, then $A(x_j,t)$ is known.\\
To determine $y_i(x_j)$, \cite{kacewicz_06} first integrates equation \ref{eqn:odes2} over $T_i:=[t_i,t_{i+1}]$:
\begin{align}
    \frac{du(x_j,t)}{dt}&=f(u(x_j,t))\\
    \int\limits_{t_i}^{t_{i+1}}du(x_j,t)&=\int\limits_{t_i}^{t_{i+1}}f(u(x_j,\tau))d\tau\label{eqn:int}\\
    u(x_j,t_{i+1})-u(x_j,t_i)&=\int\limits_{t_i}^{t_{i+1}}f(u(x_j,\tau))d\tau\\
    u(x_j,t_{i+1})&=u(x_j,t_i)+\int\limits_{t_i}^{t_{i+1}}f(u(x_j,\tau))d\tau\\
    u(x_j,t_{i+1})=u(x_j,t_i)&+\sum_{m=0}^{N_k-1}\int\limits_{t_{i,m}}^{t_{i,m+1}}f(A_{i,m}(x_j,\tau))d\tau+\sum_{m=0}^{N_k-1}\int\limits_{t_{i,m}}^{t_{i,m+1}}[f(u(x_j,\tau))-f(A_{i,m}(x_j,\tau))]d\tau
    \label{eqn:u}
\end{align}
Note that $t$ is replaced with $\tau$ in equation \ref{eqn:int} onwards to avoid confusion with the integration limits. In equation \ref{eqn:u}, $u(x_j,\tau)$ being replaced with the approximations $A_{i,m}$ results in summations to account for each of the secondary subsections. The third term on the RHS of the equation also makes the expression exact. We can: replace $u(x_j, t_i)$ with $y_i(x_j)$ since $u(x_j,t_i)\approx A_i(x_j,t_i)\equiv y_i(x_j)$, discard the third term since it is equivalent to $O(\bar{h}^{r+1})$, and rewrite $\tau=\bar{h}z$. This results in:
\begin{equation}
    y_{i+1}(x_j)=y_i(x_j)+N_k\sum_{m=0}^{N_k-1}\frac{\bar{h}}{N_k}\int_{0}^{1}f(A_{i,m}(x_j,z))dz
    \label{eqn:y_i}
\end{equation}
So each $y_i(x_j)$ can be determined by the previous $y_{i-1}(x_j)$ and the Taylor polynomials $A_{i,m}(x_j,t)$, with $y_0(x_j)$ being derived from the given initial condititon at time $t=0$. \cite{kacewicz_06} approximates the integral in equation \ref{eqn:y_i} by defining $N_k$ `knot' points $z_{m,p}$ within each secondary subinterval $T_{i,m}$, and taking the average value across the points:
\begin{equation}
    \sum_{m=0}^{N_k-1}\frac{\bar{h}}{N_k}\int_{0}^{1}f(A_{i,m}(x_j,z))dz\approx\sum_{m,p=0}^{N_k-1}\frac{\bar{h}}{N_k^2}f(A_{i,m}(x_j,z_{m,p}))
    \label{eqn:integral_approx}
\end{equation}
Unfortunately, the number of knot points becomes quite large and thus expensive to compute. To fix this problem, \cite{oz_23} introduces Chebyshev points.

\subsubsection{Chebyshev Points}
\label{subsec:chebyshevpoints}
Chebyshev points, also known as Chebyshev nodes, are the roots of the first kind of Chebyshev polynomials $T_n$, which are defined by $T_n(\cos\theta)=\cos(n\theta)$:
\begin{align}
    T_1(\cos\theta)&=\cos(1\theta)=\cos(\theta)&\Leftrightarrow T_1(x)&=x\\
    T_2(\cos\theta)&=\cos(2\theta)=2\cos^2(\theta)-1&\Leftrightarrow T_2(x)&=2x^2-1\\
    T_3(\cos\theta)&=\cos(3\theta)=4\cos^3(\theta)-3\cos(\theta)&\Leftrightarrow T_3(x)&=4x^3-3x\\
    &\hspace{1.3ex}\vdots&&\hspace{1.3ex}\vdots\nonumber\\
    T_n(\cos\theta)&=\cos(n\theta)&\Leftrightarrow T_n(x)&=2xT_{n-1}(x)-T_{n-2}(x)
\end{align}
\cite{oz_23} uses them to reduce the number of knot points from $N_k$. First, $K_{nf}$ points labelled $w_{m,p}$ are defined within each secondary subinterval $T_{i,m}$, with $p\in[0,K_{nf})$
\begin{equation}
    w_{m,p}=\frac{\cos\left(\frac{\pi p}{K_{nf}}-1\right)+1}{2}\bar{h}+t_{i,m}
\end{equation}
where we have $K_{nf}\ll N_k$. Then \cite{oz_23} define a function $q_m(g_s)$ such that $q_m(g_s)=A_{i,m}(x_j,w_{m,p})$ at $p\in[0,K_{nf})$ in the interval $T_{i,m}$:
\begin{equation}
    q_m(g_s)=\frac{q''(w_{m,p})}{w_{m,p}-w_{m,p+1}}\frac{(g_s-w_{m,p+1})^3}{6}+\frac{q''(w_{m,p+1})}{w_{m,p+1}-w_{m,p}}\frac{(g_s-w_{m,p})^3}{6}+Cg_s+D
\end{equation}
where $g_s\in T_{i,m}$ and $C,D\in\mathbb{R}$ are unknown coefficients that can be found by matching the functional values at endpoints. Finally, \cite{oz_23} introduces $K_{ns}$ knot points in the same interval $T_{i,m}$, with $K_{nf}\ll N_k\ll K_{ns}$. Substituting these points into equation \ref{eqn:integral_approx} we have:
\begin{equation}
    \sum_{m=0}^{N_k-1}\frac{\bar{h}}{N_k}\int_{0}^{1}f(A_{i,m}(x_j,z))dz\approx\frac{\bar{h}}{K_{nf}}\frac{1}{K_{ns}}\sum_{m=0}^{K_{nf}-1}\sum_{p=0}^{K_{ns}-1}f(q_m(g_s))
\end{equation}
Now, after rescaling and shifting $f$ to be in the range $[0,1]$, we can use the QAEA to evaluate the average of value $f$ on the $K_{ns}$ points.

\subsubsection{Quantum Amplitude Estimation Algorithm}
Let $\mathcal{A}$ be a unitary operator acting on a Hilbert space $\mathcal{H}$. The state obtained by applying $\mathcal{A}$ to the zero state is: $\ket{\phi}:=\mathcal{A}\ket{0}=\ket{\phi_0}+\ket{\phi_1}$. Let $\ket{n_i}:=\ket{\phi_i}/\sqrt{\braket{\phi_i|\phi_i}}$ for $i\in\{0,1\}$, and $a:=\braket{\phi_1|\phi_1}$ to rewrite the state as $\ket{\phi}=\sqrt{1-a}\ket{n_0}+\sqrt{a}\ket{n_1}$. Here, $\ket{n_1}$ is the ``good" state with $\sqrt{a}$ being its amplitude. The task of a typical QAEA is to return an estimate of $\sqrt{a}$.

The QAEA that \cite{oz_23} uses is based on a method by \cite{novak_01}. Here we define a function g to be the transformation of f after it is rescaled and shifted to be in the range [0, 1]; the QAEA aims to estimate $1/K_{ns}\sum_{i=0}^{K_{ns}-1}g(i)$. \cite{novak_01} introduces an $(n+1)$-qubit Hilbert space $\mathcal{H}'=\mathcal{H}\otimes\mathcal{H}_C$, where $\mathcal{H}$ is an $n$-qubit Hilbert space with computational basis states $\ket{i}, \: i\in\{0,\dots,N-1\}$, and $\mathcal{H}_C$ is a $1$-qubit Hilbert space with computational basis states $\ket{0}$ and $\ket{1}$. The computational basis states for $\mathcal{H}'$ are then $\ket{i}\ket{0}$ and $\ket{i}\ket{1}$. \cite{novak_01} then defines a quantum oracle $\mathcal{O}$:
\begin{align}
    \mathcal{O}\ket{i}\ket{1}&=\sqrt{g(i)}\ket{i}\ket{1}+\sqrt{1-g(i)}\ket{i}\ket{0}\\
    \mathcal{O}\ket{i}\ket{0}&=\sqrt{1-g(i)}\ket{i}\ket{1}+\sqrt{g(i)}\ket{i}\ket{0}
\end{align}
The unitary operator $\mathcal{A}$ from earlier is defined $\mathcal{A}:=\mathcal{O}(F_N\otimes I_C)$, which applies the Quantum Fourier Transform $F_N$ to the $n$ qubits and the identity operation to the remaining qubit, finally applying the oracle $\mathcal{O}$ to all $n+1$ qubits. For a state $\ket{\psi}:=\ket{0}\ket{1}$, we have:
\begin{equation}
    \ket{\psi}=A\ket{0}\ket{1}=\frac{1}{\sqrt{N}}\sum_{i=0}^{N-1}[\sqrt{g(i)}\ket{i}\ket{1}+\sqrt{1-g(i)}\ket{i}\ket{0}]
    \label{eqn:ketpsi}
\end{equation}
This makes the $\ket{n_i}$ states:
\begin{equation}
    \ket{n_0}=\frac{1}{\sqrt{N}}\sum_{i=0}^{N-1}\sqrt{\frac{1-g(i)}{\bar{g}}}\ket{i}\ket{0},\hspace{5ex}
    \ket{n_1}=\frac{1}{\sqrt{N}}\sum_{i=0}^{N-1}\sqrt{\frac{g(i)}{\bar{g}}}\ket{i}\ket{1}
    \label{eqn:n_i}
\end{equation}
where $\bar{g}$ is the desired average value. Then equation \ref{eqn:ketpsi} becomes:
\begin{equation}
    \ket{\psi}=A\ket{0}\ket{1}=\sqrt{a}\ket{n_1}+\sqrt{1-a}\ket{n_0}
\end{equation}
Here $a=\bar{g}$ as follows from equations \ref{eqn:ketpsi} and \ref{eqn:n_i}. It is then possible to use the QAEA to estimate the value of $a$ and thus $\bar{g}$.

\subsubsection{Complexity Analysis}
\cite{oz_23} shows for $n>5$ the solution error $\displaystyle\epsilon\equiv\sup_{x_j,t}|u(x_j,t)-A(x_j,t)|=O\left(\frac{1}{n^{a_k}}\right)$, where $\alpha_k:=k(q+1)-1$ and $q=r+\rho$ is the smoothness parameter from section \ref{sec:osk_temp_disc}. \cite{kacewicz_06} shows the quantum complexity as $\widetilde{O}\left((1/\epsilon)^{\frac{1}{q+1-\gamma}}\right)$, with a lower bound $\widetilde{\Omega}\left((1/\epsilon)^{\frac{1}{q+1}}\right)$ where $\gamma\in[0,1]$ is defined as an arbitrarily small parameter. \cite{gaitan_20} states that this algorithm provides a computational advantage compared to classical deterministic and randomized algorithms for non-smooth functions (where $r=0$). However, this is not the case for the 1-dimensional heat equation. In the best-case scenario we have $r=0$, $\rho\approx1$, thus $q\approx1$, and $\gamma=1$, with the result being a runtime of $\widetilde{O}(\epsilon^{-1})$. This is the same as the best quantum method for the 1-dimensional heat equation presented in \cite{linden_20} (fast r.w. amplitude estimation), and does not provide a speed-up compared to the classical Fast Fourier Transform method with a runtime of $\widetilde{O}(\epsilon^{-0.5})$.

\section{Comparative Analysis}
To make a fair comparison with classical baselines and across quantum approaches, we adopt
the LMS benchmark and require that each solver output the \emph{same} classical
scalar \(H_S(t)=\int_S u(x,t)\,dx\) to a stated additive accuracy; otherwise,
complexity claims can be misleading~\cite{linden_20}. As reviewed in
section \ref{subsec:cas-benchmark-extraction}, LMS normalize the initial mass to
\(1\), ensuring \(H_S(t)\in[0,1]\) and enabling amplitude/mean estimation from a
state that encodes the discretized solution \cite[Lemma~13]{linden_20}.

\paragraph{CAS with fair read-out.}
For the \cite{costa_22}  pipeline, preparing the (normalized) solution state costs
\(O(\kappa\log(1/\varepsilon))\)~\cite{costa_22}. Appending a bounded-output mean
estimation (Montanaro) to obtain a \emph{classical} estimate of \(H_S(t)\) to
additive error \(\varepsilon\) multiplies this by \(O(1/\varepsilon)\)
\cite[Alg.\,1; Thm.\,3; Table~1, row~1]{montanaro_17}. We allocate the error
budget as \(\varepsilon=\delta_{\text{state}}+\varepsilon_{\text{ME}}\) with, for
example, \(\delta_{\text{state}}=\varepsilon_{\text{ME}}=\varepsilon/2\), so a
single symbol \(\varepsilon\) denotes the overall additive error. Hence
\[
  T_{\text{CAS}\to\text{classical}}(\varepsilon)
  \;=\;
  O\!\Bigl(\kappa\,\frac{\log(1/\varepsilon)}{\varepsilon}\Bigr)
  \;=\;
  O\!\Bigl(m\,\frac{\log(1/\varepsilon)}{\varepsilon}\Bigr),
\]
using \(\kappa=\Theta(m)\) for the \cite{linden_20} heat-equation block system.

\paragraph{OSK (Chebyshev + QAEA).}
The Oz--San--Kara pipeline achieves
\(\widetilde{O}\!\bigl((1/\varepsilon)^{1/(q+1-\gamma)}\bigr)\), where
\(q=r+\rho\) encodes smoothness and \(\gamma\in[0,1]\) is a tuning
parameter~\cite{oz_23,gaitan_20}. In the one-dimension heat equation regime typically
considered (\(q\approx 1\), \(\gamma\to 1\)), this reduces to
\(\widetilde{O}(1/\varepsilon)\).

\paragraph{Head-to-head (one dimension, fixed grid size \(m\)).}
\begin{itemize}
  \item \cite{costa_22}  \(+\) read-out:
        \(O\!\bigl(m\,\log(1/\varepsilon)/\varepsilon\bigr)\).
  \item \cite{oz_23} (best plausible regime):
        \(\widetilde{O}(1/\varepsilon)\).
  \item Classical FFT baseline: \(\widetilde{O}(\varepsilon^{-1/2})\)
        under comparable assumptions~\cite{linden_20}.
\end{itemize}
Thus, once we require the same classical output \(H_S(t)\), both quantum options
exhibit the familiar \(1/\varepsilon\) precision overhead (\cite{costa_22}  differs by a mild
\(\log(1/\varepsilon)\) and a linear-in-\(m\) prefactor from the linear-system
structure), and neither outperforms the FFT baseline on this one-dimensional benchmark.
\cite{costa_22}'s strengths with optimal \(\log(1/\varepsilon)\) dependence in state
preparation and transparent \(\kappa\)-sensitivity are neutralized at the
comparison point by the necessary classical extraction.

\paragraph{Higher-dimensional outlook (hypothesis).}
Extending \cite{costa_22}  to \(d>1\) preserves sparsity but typically changes conditioning so
that \(\kappa=\Theta(m^{2/d})\) for common discretizations, suggesting a
state-preparation cost \(O(m^{2/d}\log(1/\varepsilon))\); with the same
Option-A read-out, this yields
\[
  O\!\Bigl(m^{2/d}\,\frac{\log(1/\varepsilon)}{\varepsilon}\Bigr).
\]
Actual performance depends on the adiabatic gap and data-loading overheads, which
are problem dependent and deserve a dedicated analysis. \cite{oz_23} generalizes to higher
dimensions as well, but its precision exponent remains \(\ge 1\) in the
physically relevant regimes, so classical multigrid/FFT remain strong baselines.

\paragraph{Conclusion.}
Under the LMS benchmark (normalized total heat in a region), \cite{costa_22} provides a
principled QLSP route with clean dependence on \(\kappa\) and \(\varepsilon\),
but once we fairly convert the quantum state to the required classical number,
its precision scaling aligns with other quantum pipelines and does not beat the
FFT baseline in one dimension. Practical viability hinges on constants and implementation
details---notably state preparation and the adiabatic gap (for \cite{costa_22} ), and
pre-/post-processing plus oracle design (for \cite{oz_23} ). Because full field recovery
would require quantum tomography with at least linear dependence on the
Hilbert-space dimension (exponential in qubits), focusing on bounded observables
like \(H_S(t)\) is essential for fair and realistic comparisons~\cite{apeldoorn_22}.
Future gains may come from observable-aware linear-system solvers, variance
reduction in read-out, or hybrid discretizations that reduce \(\kappa\) without
compromising sparsity.

\section*{Acknowledgements}
\addcontentsline{toc}{section}{Acknowledgements}
We thank Daochen Wang for taking the time to review our work and offer valuable feedback and guidance on the manuscript.

\bibliography{references}
\addcontentsline{toc}{section}{References}
\bibliographystyle{alphaurl}
\end{document}